\documentclass [preprint,aps,showpacs]{revtex4}
\usepackage[final]{graphics}
\usepackage{amssymb}
\usepackage{amsfonts}
\usepackage{epsfig}
\usepackage{graphicx}

\begin{document}

\title{Contact process with mobile disorder}

\author{Ronald Dickman\footnote{dickman@fisica.ufmg.br}
}
\address{Departamento de F\'{\i}sica, Instituto de Ci\^encias Exatas,\\
Universidade Federal de Minas Gerais \\
C. P. 702, 30123-970, Belo Horizonte, Minas Gerais - Brazil }

\date{\today}

\begin{abstract}

I study the absorbing-state phase transition in
the one-dimensional contact process with mobile disorder.
In this model the dilution sites, though permanently inactive,
diffuse freely, exchanging positions with the other sites,
which host a basic contact process.
Even though the disorder variables are not quenched, the critical
behavior is affected: the critical exponents $\delta$ and $z$, the
ratio $\beta/\nu_\perp$ and the moment ratio $m= \langle \rho^2 \rangle/\rho^2$
take values different from those of directed percolation, and appear
to vary with the vacancy diffusion rate.  While the survival probability starting from a
single active seed follows the usual scaling, $P(t) \sim t^{-\delta}$, at the
critical point, the mean number of active sites and mean-square spread grow
more slowly than power laws.  The critical creation rate increases
with the vacancy density $v$ and diverges at a value $v_c < 1$.
The scaling behavior at this point appears to be simpler than for smaller
vacancy densities.

\end{abstract}


\maketitle

\section{Introduction}

Phase transitions to an absorbing state continue
to attract much attention in nonequilibrium statistical
physics \cite{marro,hinrichsen,granada,lubeck04,odor04},
and have stimulated
efforts to characterize the associated universality classes.
Absorbing-state
transitions arise in models of epidemics, population dynamics and
autocatalytic chemical reactions, and underly self-organized criticality
in sandpile models.
The simplest models exhibiting absorbing-state phase transitions
are the contact process (CP) \cite{harris,marro} and its
discrete-time version, directed percolation (DP).

Scaling at absorbing state phase transitions has been studied
extensively, both theoretically and numerically, and more recently, in
experiments \cite{takeuchi}.  A central conclusion is that
critical behavior of the DP type is generic for
models exhibiting a continuous phase transition to an absorbing state, in the
absence of any additional symmetries or conserved quantities \cite{janssen,grassberger}.
In particular, allowing particles to diffuse at a finite rate does not
affect the critical behavior of the contact process \cite{iwandcp}.

The effect of quenched disorder on the CP, in the form of random dilution,
as well as of randomly varying creation rates associated with each site,
has been investigated in detail, beginning with the studies of Noest \cite{noest},
and in subsequent analyses of its unusual scaling
properties \cite{agrd,janssen97,bramson,webman,cafiero,hooyberghs,vojta,hoyos,deoliveira08}.
Little attention has been given to the effect of {\it mobile} disorder, for example
in the form of diffusing passive sites.
An exception is the recent study by Evron et al. of a CP in which certain sites
have a higher creation rate than others, with the ``good" and ``bad" sites
diffusing on the lattice \cite{evron}.
These authors report anomalies in the off-critical behavior of the model.

Lack of interest in the effect of mobile disorder at an absorbing-state
phase transition presumably reflects a belief that it should
not lead to any significant changes in scaling properties, and might in fact be
irrelevant.  In equilibrium, {\it annealed} disorder at fixed concentration
leads to ``Fisher renormalization"
of the critical exponents if the specific heat exponent $\alpha > 0$ in the pure model, and does not
alter the exponents if $\alpha < 0$ \cite{fisherren}.
The condition for Fisher renormalization (FR), like that of Harris' criterion \cite{harris74}, can be
extended to models for which (as in the present case) $\alpha$ is not defined \cite{berker}, by using the
equivalent relation, $d \nu < 2$.  (In the present context $\nu$ corresponds to the
correlation length exponent $\nu_\perp$.)  A subtlety nevertheless arises in
extending the notion of FR to nonequilibrium systems, because this phenomenon
requires {\it equilibrium} between the degrees of freedom associated with the
disorder (such as vacancy positions) and the other elements comprising
the system (e.g., spin variables).  In the present case there is no energy
function that could favor (via a Boltzmann factor $e^{-\beta H}$) one disorder configuration
over another; all disorder configurations equally probable, since the CP
degrees of freedom have no influence over the vacancy dynamics.  We therefore speak of
``mobile" rather than ``annealed" disorder.
(Of course, those disorder configurations that are more favorable to the maintenance
of activity in the CP will survive longer and so might dominate long-time averages
over {\it surviving} realizations.)

In the FR scenario, the exponent $\alpha$ is replaced by $-\alpha/(1-\alpha)$, whilst the
other static critical exponents (with the exception of $\eta$) are multiplied by a factor
of $1/(1-\alpha)$.  Annealed disorder is expected to be relevant for critical
{\it dynamics} if $\nu (d+z) < 2$, where $z$ is the dynamic exponent \cite{alonso-munoz}.
In the case of the contact process, under the FR scenario, the critical exponents $\beta$ and
$\nu_\perp$ would be renormalized by the same factor.
(If we confide in the hyperscaling relation
$\alpha = 2 - d\nu_\perp$, we would expect that $\beta \to \beta/(d\nu_\perp - 1)$,
and similarly for the other static exponents.)
It is less clear what would happen
to the exponents governing critical dynamics, such as $z$ or $\delta$, or
to an order-parameter moment ratio such as
$m \equiv \langle \rho^2 \rangle/\langle \rho \rangle^2$.

In this work I study the CP with mobile vacancies.  The vacancies diffuse freely,
at rate $D$, exchanging positions with nondiluted sites, be they active or inactive,
which host the usual CP.  Aside from its
intrinsic interest, processes of this sort should find application in epidemic modelling
(wandering of immune individuals in a susceptible population), and in population dynamics
(for example, a diffusing population of a third species in a predator-prey system).  The
presence of diffusing impurities is in principle relevant to experimental realizations
of directed percolation in fluid media \cite{takeuchi}.
It transpires that the critical creation rate $\lambda_c$ diverges at a certain
vacancy density $v_c(D) < 1$, so that an active state is impossible for
$v > v_c$.  It is thus of interest to study the limiting case of $\lambda \to \infty$,
which corresponds the limit of an epidemic that propagates extremely rapidly, and can
be impeded only by fragmenting the population.

The remainder of this paper is organized as follows.  In Sec. II I define the model and discuss
some of its basic properties as well as the results of the pair approximation.
Sec. III presents the results of various simulation approaches, and in Sec. IV
the implications for scaling behavior are discussed.

\section{Model}

\begin{figure}[h]
\epsfysize=10cm \epsfxsize=11cm \centerline{ \epsfbox{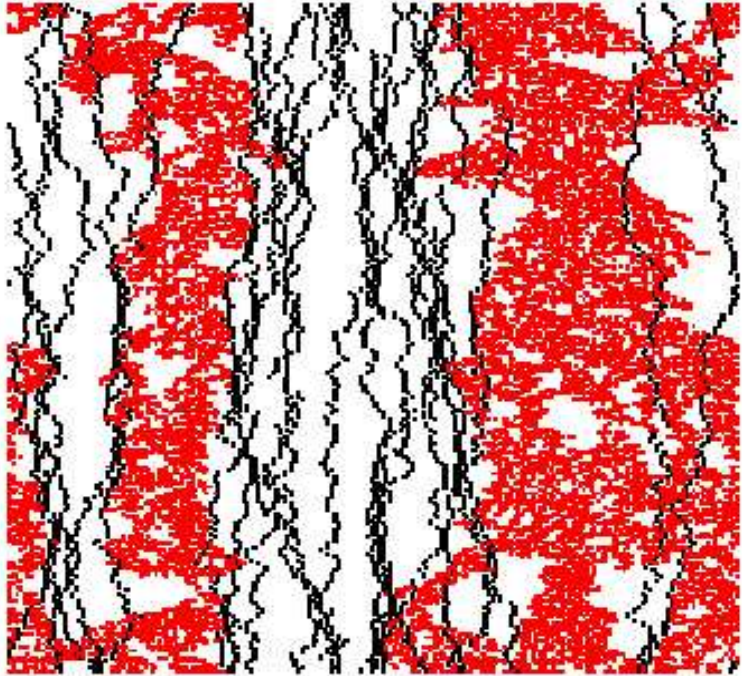}}
\caption{\footnotesize{Typical space-time evolution, $L=200$, $v=0.1$, $D=1$,
and $\lambda = 4.095$.  Red points: active sites; black points: vacancies.
This image corresponds to 200 time steps, with time increasing downward.}}
\label{cpmbvevb}
\end{figure}

In the basic contact process\cite{harris} each site of a lattice is in one of two states,
``active" or ``inactive".  Transitions of an active site to the inactive
state occur spontaneously (i.e., independent of the other states of the sites)
at unit rate, while an inactive site becomes active at rate $\lambda n/q$,
where $n$ is the number of active nearest neighbors and $q$ the total number of
neighbors.  The configuration with all sites inactive is absorbing.
The model is known to exhibit a continuous phase transition
between the active and absorbing states at a critical creation rate $\lambda_c$;
in the one-dimensional CP $\lambda_c = 3.29785(2)$.  (Figures in parentheses denote uncertainties.)
The transition falls in the universality class of directed percolation.
In the CP with mobile disorder (CPMV), a fraction $v$ of the sites are {\it vacant},
and do not participate in the usual CP dynamics.  The vacancies hop on the lattice
at rate $D$.  In a hopping transition, a vacancy trades its position with
its right or left neighbor.  Denoting vacancies, and active and inactive sites
by v, 1 and 0, respectively, we have the transitions v1 $\to$ 1v and v0 $\to$ 0v
at rate $D/2$, and similarly for vacancy hopping to the left.
(From here on, an ``inactive" site denotes one that is not a vacancy, but that
happens not to be active.)
The typical evolution shown in Fig.~\ref{cpmbvevb} illustrates that regions
of higher than average vacancy density tend to be devoid of activity, and vice-versa.

The CPMV is related to a model studied recently by Evron, Kessler and Shnerb (EKS) \cite{evron}, in which half
the sites are ``good", meaning they have a creation rate $\lambda_1$, while the
creation rate at the other, ``bad" sites is $\lambda_2 < \lambda_1$.  The good and bad sites
diffuse on the lattice at rate $D$, independent of their state (active or inactive).
The CPMV resembles the limit $\lambda_2 \to 0$ of the EKS model, but the correspondence is not exact,
since bad sites can in fact become active in the latter system, even if $\lambda_2 = 0$.
(For $\lambda_2 =0$, a bad site can never produce offspring, but it can always be activated
by an active neighbor.  In the CPMV by contrast, the vacancies can never be activated.)

While the EKS model with $\lambda_2 > 0$ represents a weaker form of disorder than the CPMV,
one might expect the two models to fall in the same universality class.
In principle, both models should have the same continuum limit, given by the following
pair of stochastic partial differential equations:

\begin{equation}
\frac{\partial \rho}{\partial t} = {\cal D} \nabla^2 \rho  + (a + \gamma \phi)\rho
- b \rho^2 + \eta({\bf x},t)
\label{ft1}
\end{equation}
and
\begin{equation}
\frac{\partial \phi}{\partial t} = \overline{\cal D} \nabla^2 \phi + \nabla \cdot {\boldmath \xi}({\bf x},t)
\label{ft2}
\end{equation}

\noindent where $\rho({\bf x},t)$ is the activity density and $\phi({\bf x},t)$ the density of nonvacant
(or ``good") sites, whose evolution is completely independent of $\rho$.
Without the term $\propto \phi \rho$, Eq. (\ref{ft1}) is the usual continuum
description of directed percolation \cite{janssen,grassberger,cardy}, with $\eta({\bf x},t)$ a zero-mean,
Gaussian noise with autocorrelation
$\langle \eta({\bf x},t)\eta({\bf y},s) \rangle \propto \rho({\bf x},t) \delta({\bf x} - {\bf y}) \delta(t-t')$.
The second term on the r.h.s. of Eq. (\ref{ft2}) similarly represents a conserved delta-correlated noise.
It is not my intention to analyze these equations here.  They merely serve to highlight the
observation that if, as seems plausible, they represent the continuum limit of both models, then
the models should have a common scaling behavior.

Suppose that, as seems reasonable, the CPMV exhibits a continuous phase transition to the
absorbing state at a critical creation rate $\lambda_c(v,D)$.
For any nonzero vacancy concentration, $\lambda_c$ must diverge as $D \to 0$,
since we are then dealing with a CP on the line with frozen dilution, which has
no active phase for finite $\lambda$.  The limiting value of $\lambda_c$ as
$D \to \infty$ is a subtler issue.  If all three species
were to diffuse infinitely rapidly, the system would be well mixed, and would
exhibit mean-field like behavior.
Note however that {\it vacancy} diffusion does not change
the ordering of the active and inactive sites.  Thus infinitely rapid
vacancy diffusion does not correspond to a mean-field limit, but rather
to a system in which a fraction $v$ of
sites are momentarily replaced, independently at each
instant and at each lattice site, by inert elements.  Hence we should expect to observe
the behavior of the {\it pure model}, at an effective reproduction rate $\lambda' = (1-v) \lambda$,
so that the critical point is renormalized to $\lambda_c' = \lambda_{c,pure}/(1-v)$ in this
limit.

While disorder that diffuses infinitely rapidly is irrelevant to critical
behavior, one may argue that it is relevant for finite $D$.  Consider a correlated region in
the CP, with characteristic size $\xi$ and duration $\tau$.  If fluctuations in the vacancy density
on this spatial scale relax on a time scale $\tau_\phi \ll \tau$, then the CP will be subject, effectively,
to a disorder that is uncorrelated in time, which is irrelevant\cite{hinrichsenbjp}.
But since the fluctuations
relax via diffusion, $\tau_\phi \sim \xi^2$.  In the neighborhood of the critical point,
$\xi \sim |\lambda - \lambda_c|^{-\nu_\perp}$ and $\tau \sim \xi^z$, with $z = \nu_{||}/\nu_\perp$,
so that $\tau_\phi \sim \tau^{2/z}$,
suggesting that diffusing disorder is relevant for $z < 2$, provided of course
that {\it quenched} disorder is relevant.  (The two conditions are conveniently
written as $d\nu_\perp < 2$ and $d \nu_{||} < 4$.)  In directed percolation this inequality
is satisfied in $d < 4$ space dimensions.

It is perhaps worth mentioning that the present model is not equivalent to the diffusive epidemic
process (DEP) \cite{dep}.  In the latter model, there are two kinds of particle (healthy and
infected, say), which diffuse on the lattice. Infected particles recover spontaneously while a healthy
particle may become infected if it occupies the same site as an infected particle.
(There is no limit on the occupation number at a given site in the DEP.)  While it is tempting to identify
the active and inactive sites of the CPMV with (respectively) the infected and healthy particles of the DEP,
the analogy is not valid since, in the CPMV, vacancy diffusion does not
cause the nondiluted sites to change their relative positions, as noted above.

\begin{figure}[h]
\epsfysize=10cm \epsfxsize=12cm \centerline{ \epsfbox{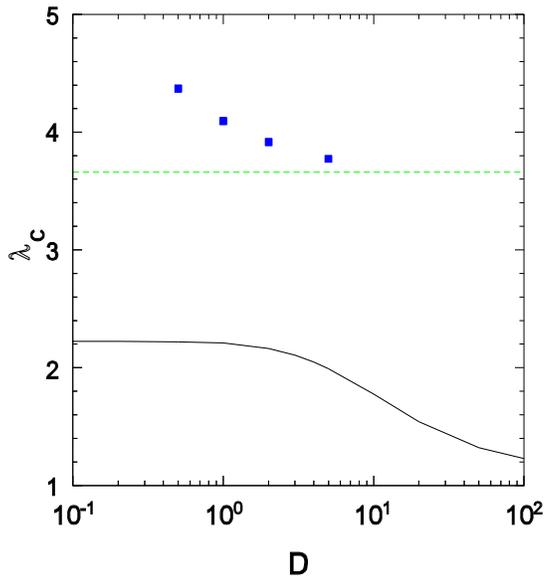}}
\caption{\footnotesize{Critical reproduction rate $\lambda_c$
versus diffusion rate $D$, for dilution $v=0.1$.  Points: simulation;
solid line: pair approximation; dashed line: limiting ($D \to \infty$) value,
$\lambda_c = \lambda_{c,pure}/v$.
}}
\label{lcv01}
\end{figure}

A simple and often qualitatively reliable approach to estimating the phase diagram is via
$n$-site approximations, where $n=1$ corresponds to simple mean-field theory. At order $n$,
the $m$-site joint probability distributions (for $m > n$) are approximated on the basis of
the $n$-site distribution \cite{benav,mancam}.  Since simple mean-field theory is insensitive to diffusion,
the lowest order of interest here is $n=2$ (the pair approximation), which is
derived in the Appendix.  The resulting transition line, $\lambda_c (D)$, is compared
with simulation results (for $v=0.1$) in Fig. \ref{lcv01}.
The pair approximation turns out to be in rather poor agreement
with simulation.  In particular, it fails to show
that $\lambda_c$ diverges as $D \to 0$.  Some improvement might be expected using larger clusters, but
it is not clear if the correct asymptotic behavior would be captured even for
large $n$.

For a fixed diffusion rate $D>0$, we expect $\lambda_c$ to increase monotonically with
vacancy concentration $v$.  One may ask whether $\lambda_c$ remains finite for
any $v < 1$, or whether it diverges at some vacancy concentration that is strictly
less than unity.  Now, for very large $\lambda$, a string of two or more active
sites is essentially immortal, for each time a site becomes inactive (which occurs at rate 1),
it is immediately reactivated by its neighbor.  Thus only isolated (nondiluted) sites can
remain inactive.  For $v$ sufficiently large, we expect an isolated active site
to become inactive before it can make contact with another nondiluted site, so that
$\lambda_c \to \infty$ at some critical vacancy concentration $v_c(D) < 1$.  The following simple
argument furnishes an estimate of $v_c$.  The rate of loss of activity due to
isolated active sites becoming inactive is $ \simeq \rho v^2$ (that is, the mean-field
estimate for the density of isolated active sites).
On the other hand, the rate of reactivation of inactive sites may be estimated as
$D \rho v (1 - \rho - v)$, i.e., the vacancy diffusion rate times the mean-field estimate
for the density of active-vacant-inactive triplets.  The resulting equation of motion,
$d \rho / dt = - v^2 \rho + D \rho v (1 - \rho - v)$, has the stationary solution
$\rho = 1 - v - v/D$, so that the activity density vanishes at $v = D/(1+D) \equiv v_c$.
While this approximation is too crude to yield a quantitative prediction, the
expression for $v_c$ does at least attain the correct limiting values for $D \to 0$
and $D \to \infty$.  (The pair approximation does not yield a useful
prediction for $v_c$.) Simulation results for $v_c$ are reported in Sec. IIId.

\section{Simulations}

I performed three series of simulations of the model.
The first series follows the evolution starting from a maximally occupied
configuration (all nonvacant sites occupied) from the initial transient
through the quasi-stationary (QS) state, in which mean properties of surviving
realizations are time-independent.  These simulations will be referred to
as ``conventional", to distinguish them from the second series, which employ the QS simulation
method \cite{qssim}. The third series studies the spread of activity
starting from an initial seed.  In all cases, the initial positions of the $vL$ vacancies
are selected at random from the $L$ sites of the ring.  A new selection is
performed at each realization.

\subsection{Conventional simulations}

In conventional simulations (CS), the fraction $\rho$ of active sites and
the moment ratio $m=  \langle \rho^2 \rangle/ \rho^2$ are calculated as functions
of time.  The time increment associated with a given event is
$\Delta t = [(1+\lambda) N_a + D N_v]^{-1}$,
where $N_a$ and $N_v$ denote the numbers of active sites and of vacancies, respectively.
$\rho(t)$ and $m(t)$ are calculated as averages over surviving realizations in time intervals
that for large $t$ represent uniform intervals of $\ln t$, a procedure sometimes
called ``logarithmic binning".  In addition to the QS values of $\rho$ and $m$,
conventional simulations furnish the mean lifetime $\tau$ from the
asymptotic decay of the survival probability,
$P_s \sim e^{-t/\tau}$. Estimates for the critical
exponents $\delta$ and $z$ are obtained using the relations
$\rho(t) \sim t^{-\delta} $ and $m(t)-1 \sim t^{1/z}$ \cite{silva}.
The latter relations describe the approach to the QS regime in a large system at its
critical point.  Finite-size scaling (FSS) theory implies that, at the critical point,
$\rho_{QS} \sim L^{-\beta/\nu_\perp}$, while $\tau \sim L^z$, and that
$m$ converges to a finite limiting value as $L \to \infty$ \cite{dic-jaff}.
Note that we have, in principal, two independent ways to estimate the exponent $z$,
one involving the scaling of the lifetime with system size,
the other using the approach of $m$ to its QS value.
(These scaling behaviors have been verified at absorbing-state transitions in various models
without disorder;
scaling in the presence of quenched disorder is generally more complicated.
Whether simple FSS applies for mobile disorder is an open question; the results shown below
provide partial confirmation.)

\begin{figure}[h]
\epsfysize=9cm \epsfxsize=12cm \centerline{ \epsfbox{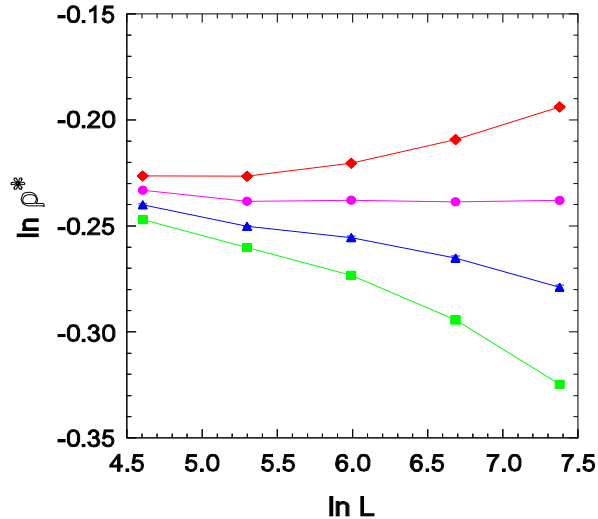}}
\caption{\footnotesize{Scaled order parameter $\rho^* = L^{\beta/\nu_\perp} \rho$
versus system size in conventional simulations for
$v=0.1$, $D=2$, and (lower to upper) $\lambda$ = 3.905, 3.910, 3.915, and 3.920.
Error bars are smaller than symbols.}}
\label{rhosb}
\end{figure}

\begin{figure}[h]
\epsfysize=9cm \epsfxsize=12cm \centerline{ \epsfbox{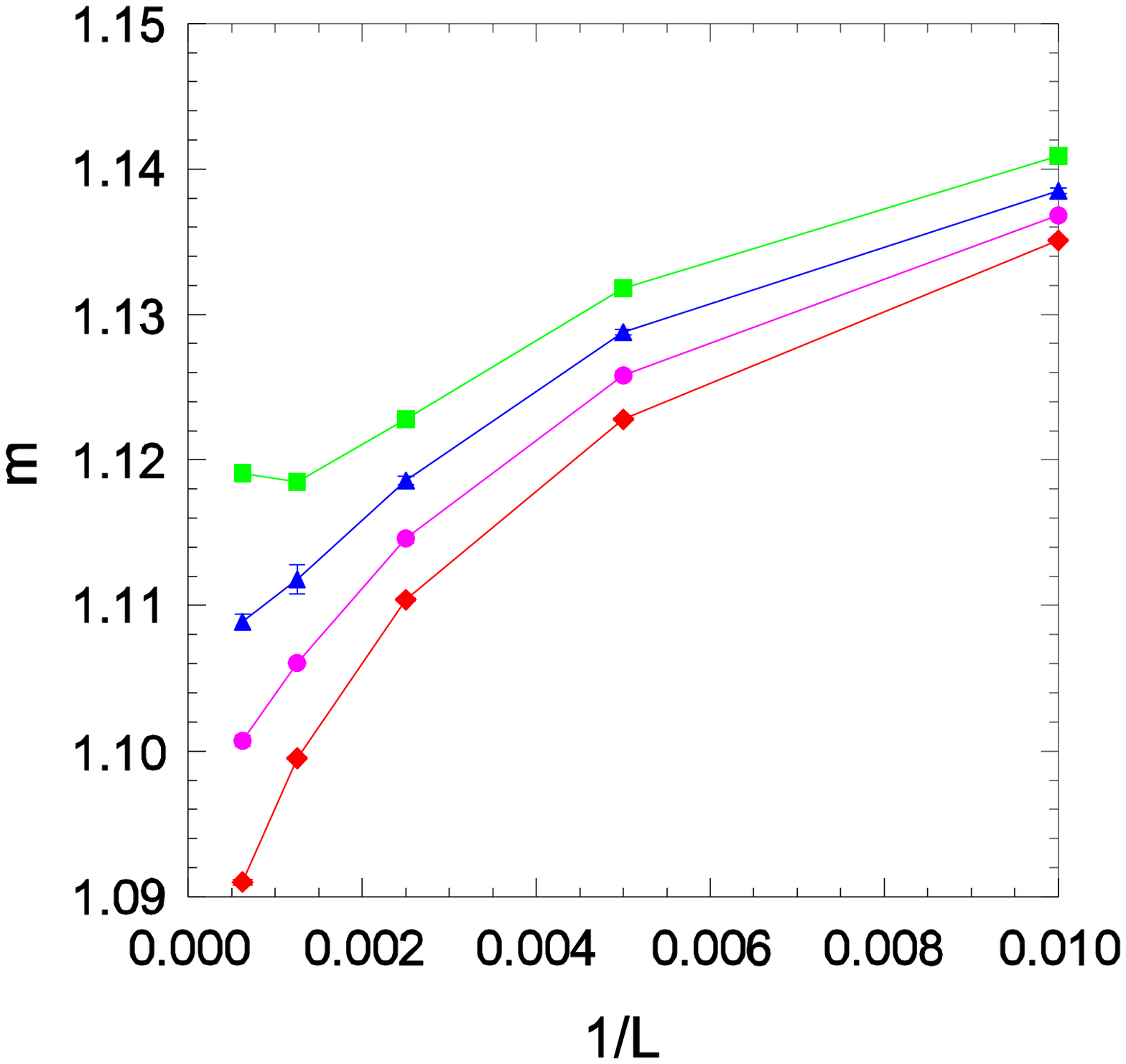}}
\caption{\footnotesize{Moment ratio $m=  \langle \rho^2 \rangle/ \rho^2$
versus inverse system size in conventional simulations for
$v=0.1$, $D=2$, and (upper to lower) $\lambda$ = 3.905, 3.910, 3.915, and 3.920.
}}
\label{mb}
\end{figure}

I performed CS for system sizes $L=100, 200,...,1600$,
averaging over $N_r$ independent realizations, where $N_r = 100\,000$ for $L=100$, and
decreases gradually with system size, to a value of 2$\,$000 - 5$\,$000 for $L=1600$.
The run times are such that all realizations eventually fall into the absorbing state.
In analyzing the CS data, I use two criteria for determining the critical reproduction rate
$\lambda_c$: (1) power-law scaling of the QS density, $\rho_{QS}$, with system size,
and (2) approach of the moment ratio $m_{QS}$ to a well defined limiting value with increasing
system size.  Verification of (1) is facilitated by studying the curvature of $\ln \rho_{QS}$ versus
$\ln L$, by calculating a quadratic fit to the data, and determining the range of $\lambda$ for
which the quadratic term is zero to within uncertainty.  It is also useful to plot
$L^{\beta/\nu_\perp} \rho_{QS}$ versus $L$, on log scales, to check for curvature visually,
as illustrated in Fig. \ref{rhosb}.
(Here $\beta/\nu_\perp$ is estimated from the data for the
value of $\lambda$ showing the least curvature.)
Verification of the second criterion is facilitated by plotting
$m$ versus $1/L$; for subcritical values this graph veers upward as $1/L \to 0$ and vice-versa,
as shown in Fig. \ref{mb}.

\begin{figure}[h]
\epsfysize=9cm \epsfxsize=12cm \centerline{ \epsfbox{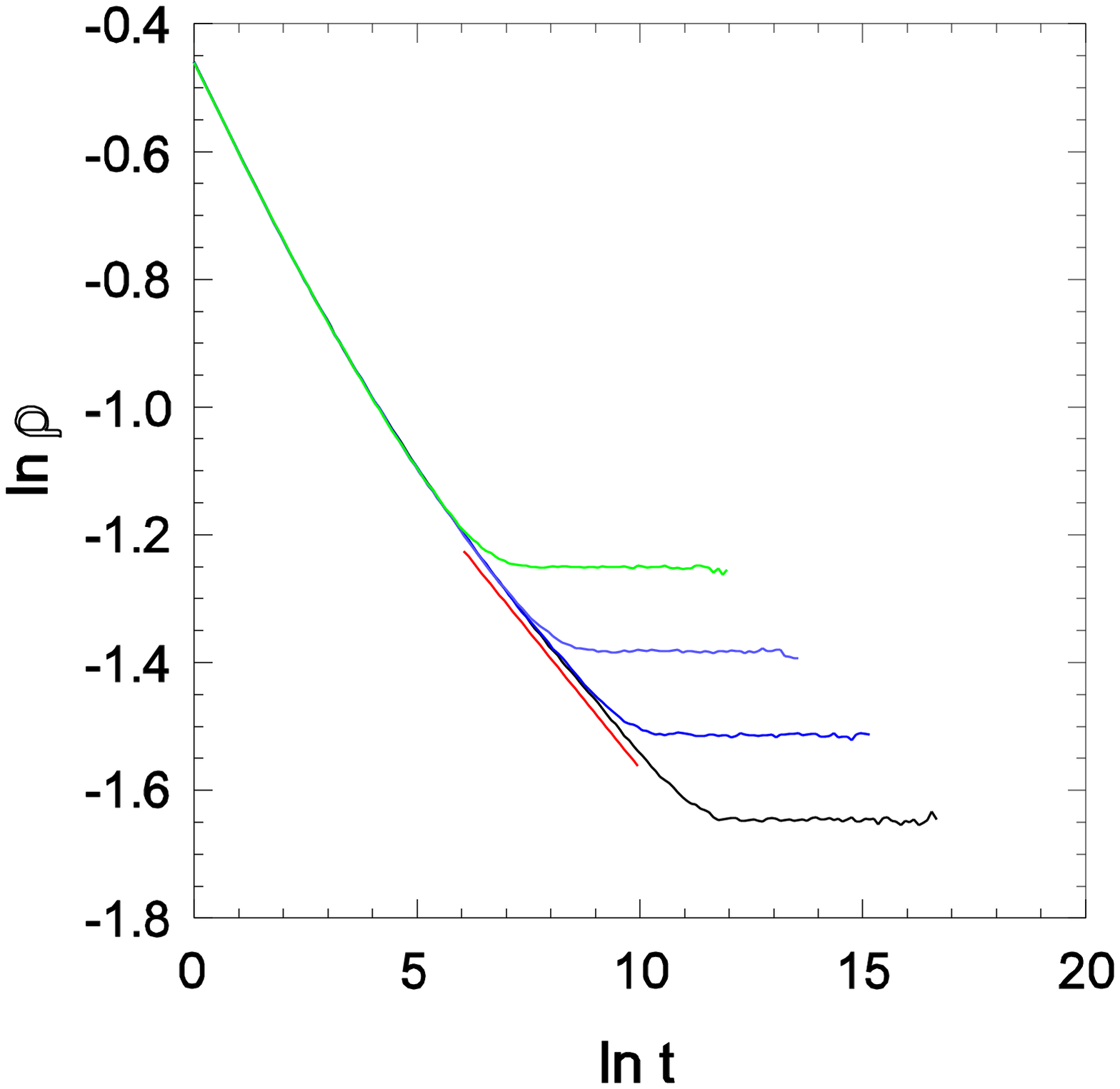}}
\caption{\footnotesize{Order parameter $\rho(t)$ for
$v=0.1$, $D=1$, and $\lambda = 4.099$. System sizes (upper to lower)
$L$ = 200, 400, 800, and 1600.  The slope of the straight line is -0.0863.}}
\label{rt099}
\end{figure}

\begin{figure}[h]
\epsfysize=9cm \epsfxsize=12cm \centerline{ \epsfbox{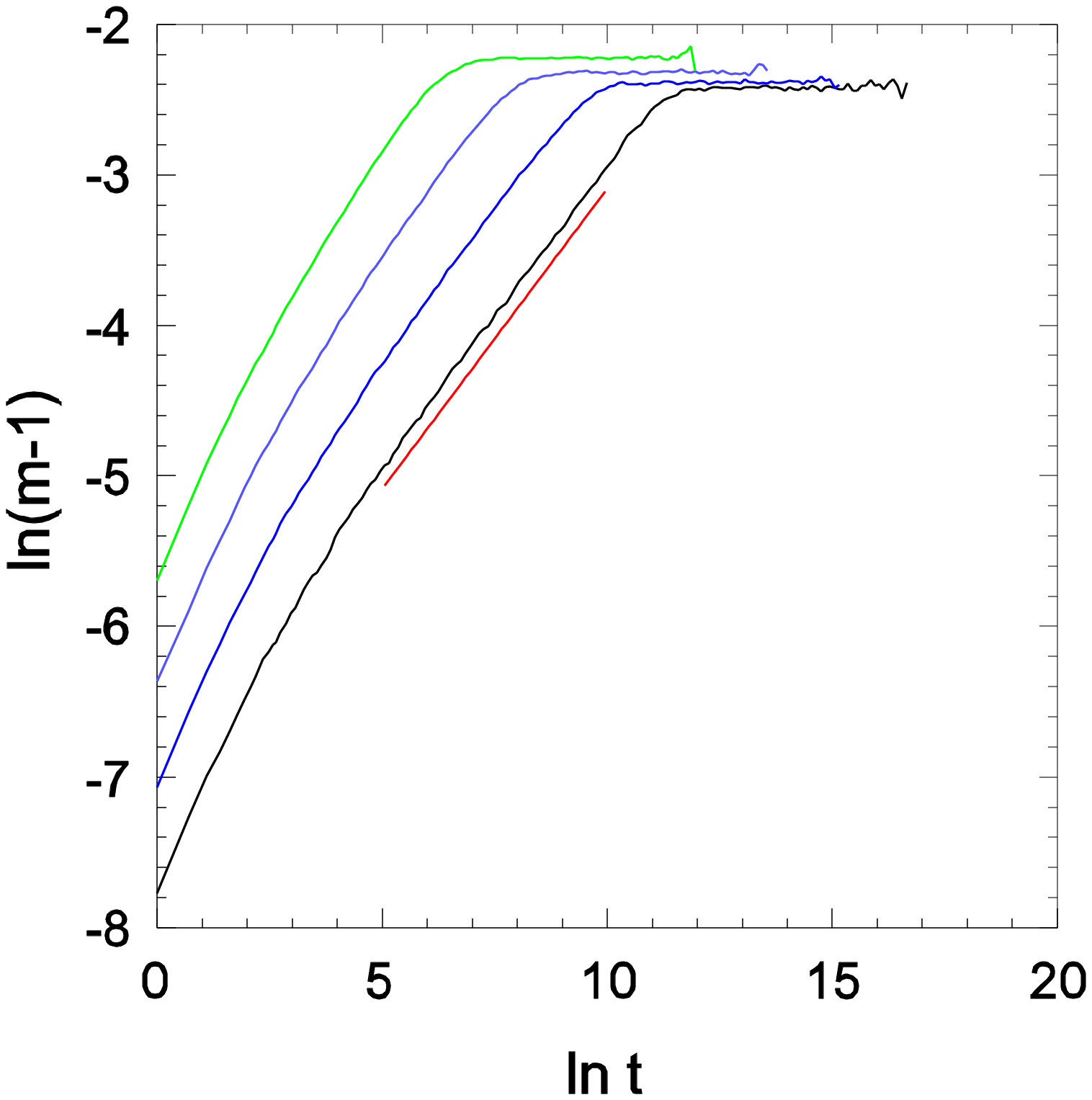}}
\caption{\footnotesize{Evolution of $m(t)-1$ for
$v=0.1$, $D=1$, and $\lambda = 4.099$. System sizes as in Fig. \ref{rt099}.
The slope of the straight line is 0.40.}}
\label{mmt099}
\end{figure}

The behavior of the activity density as a function of time, at criticality,
is shown in Fig. \ref{rt099}, for the case $v=0.1$ and $D=1$.  In the pure contact process
the activity follows a power law, with very weak corrections to scaling, for comparable
times and system sizes \cite{marro}.  Here by contrast, there is a significant
positive curvature to the plots of
$\rho(t)$ on log scales.  It is nevertheless possible to estimate the exponent $\delta$
from the data at longer times and larger system sizes.  Since the graphs of $\rho(t)$ are not
linear (on log scales) it is not possible to collapse the data onto a unique master curve describing
both the stationary and transient regimes, as is in fact possible in the contact process
without dilution.
The quantity $m(t)-1$ shows cleaner power-law scaling over a somewhat larger period, as shown in
Fig. \ref{mmt099}, allowing a rather precise estimate for the exponent $z$.

\begin{figure}[h]
\epsfysize=9cm \epsfxsize=12cm \centerline{ \epsfbox{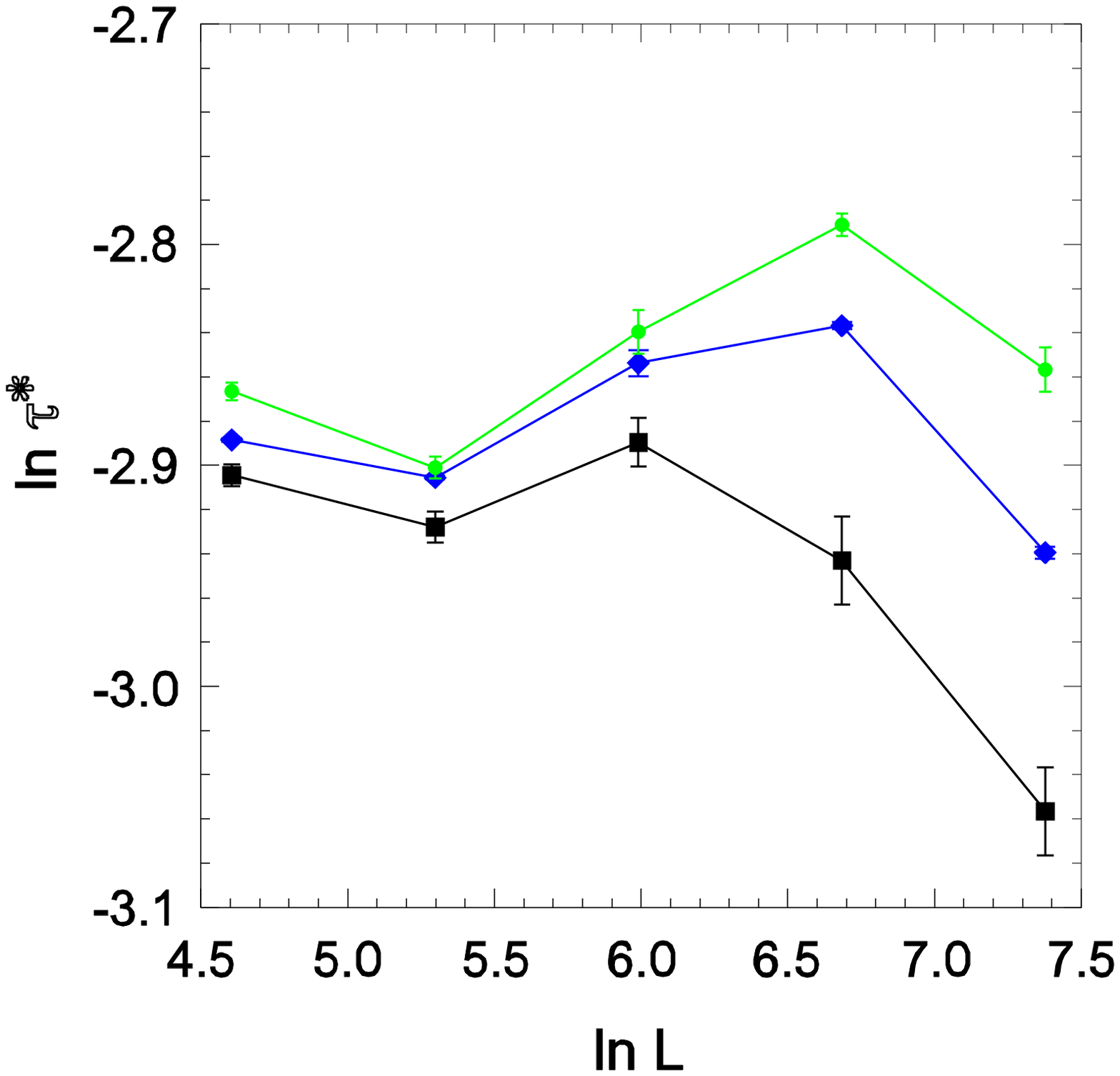}}
\caption{\footnotesize{Scaled lifetime $\tau^* = L^{-z} \tau$ versus system size
for $v=0.1$, $D=1$, and (lower to upper) $\lambda = 4.097$, 4.099, and 4.101.}}
\label{tausc}
\end{figure}

Table I summarizes the results of the conventional simulations.
The critical parameters such as $\beta/\nu_\perp$ and $m$
are calculated for each value of $\lambda$ simulated, and then evaluated at $\lambda_c$ using
linear interpolation.  Each fit to the data involves an error bar, to which we must add the (typically
larger) uncertainty induced by the uncertainty in $\lambda_c$ itself.
In Table I, the values cited for the dynamic exponent $z$ are obtained from the growth of the
moment ratio (using $m-1 \sim t^{1/z}$), and not from the FSS relation for
the lifetime, $\tau \sim L^z$.  In fact, the growth of the lifetime appears to
be slower than a power law at the critical point, as shown in Fig. \ref{tausc} for $v=0.1$ and
$D=1$.  In all the cases shown (including $\lambda = 4.101$, above the critical value),
the plots of $\tau^* = L^{-z} \tau$ curve downward.  (In this plot, for purposes of
visualization I use $z=2.4$, derived
from a linear fit to the data form $\tau (L)$ at $\lambda = 4.099$.)

As noted in the Introduction, averages restricted to surviving realizations may
yield statistical properties of the vacancies that differ from those of a random
mixture.  I studied the density of nearest-neighbor vacancy pairs; for a
random mixture $\rho_{vv} = N(N-1)/L^2$, where $N=vL$ is the number of vacancies.
At the critical point ($\lambda = 4.099$, $D=1$, and $v=0.1$),
I find that $\rho_{vv}$ grows with time, reaching a value about 2\% greater than
the random mixture value in a system with $L=100$.  For system sizes $L=200$ and 800,
the increase over the random mixture value is 1\% and 0.2\%, respectively.
Thus selection effects on the vacancy configuration
appear to be quite weak.

\subsection{Quasistationary simulations}

This simulation method is designed to sample directly the quasistationary
(QS) regime of a system with an absorbing state, as discussed in detail in
\cite{qssim}.  The method involves maintaining a list of active configurations
during the evolution of a given realization.  When a transition to the
absorbing state is imminent, the system is instead placed in one of the active
configurations, chosen randomly from the list.  This procedure has been shown to
reproduce faithfully the properties obtained (at much greater computational
expense) in conventional simulations.
The theoretical basis for the QS simulation method remains
valid in the presence of variables that continue to evolve even in the
absence of activity, such as the vacancy positions in the CPMV, or
the positions of healthy particles in the diffusive epidemic process \cite{epidif2}.

I performed QS simulations on rings of
$L$ sites, as the conventional
simulations. The simulation yields a
histogram of the accumulated time during which the system has exactly 1, 2,...n,...,
active sites, which is used to calculate the QS order parameter $\rho_{QS}$
and moment ratio $m_{QS}$.  The lifetime
$\tau$ is given by the mean time between attempted visits to the absorbing state,
in the QS regime.
Results are obtained from averages over 10 independent realizations, each
lasting $t_{max} = 10^8$ time units, with an initial portion (10$^4$ to $ 10^5$ time steps,
depending on the system size), discarded to ensure the system has attained the QS state.
One thousand saved configurations are used to implement the QS simulation procedure \cite{qssim}.
Values of the replacement probability $p_{rep}$
range from $10^{-3}$ to $6 \times 10^{-5}$  (smaller values for larger $L$); during the initial
relaxation period, $p_{rep}=0.01$.

Comparing the results of QS and conventional simulations near the critical point,
I find no significant differences for $v=0.1$ and $D=0.5$, in systems of up to 1600 sites.
For $D=1$, the results for $\rho$ and $m$ again agree to within uncertainty,
but the QS simulations yield slightly smaller values for the lifetime than
conventional simulations.  (For $L=800$, the difference is about 5\%.)
The difference can be understood if we suppose that
certain initial vacancy configurations are more favorable than others for the survival
of activity.  In the conventional simulations
each initial configuration makes a single contribution to the average survival time, whereas
in QS simulations, realizations starting with less favorable configurations (with shorter lifetimes) visit the
absorbing state more frequently than those starting from favorable ones, and so tend to
dominate the average for the lifetime.
This presumably reflects the slow relaxation of the vacancy density fluctuations.
An advantage of conventional simulations in this respect is that many more initial
configurations are sampled than in the QS simulations.  (The former also
provide information on the relaxation process, that is not available in QS
simulations.)
Despite the minor difference in lifetime estimates, the otherwise excellent agreement
between the two simulation methods lends confidence to the results obtained.

\subsection{Spreading simulations}

In light of the surprising results of the conventional simulations, it is of interest
to study the model using a complementary approach, which follows the spread of activity,
starting from a localized seed.  Such spreading or ``time-dependent" simulations were
pioneered in the context of contact process by Grassberger and de la Torre \cite{torre}.
Starting with a single active site at the origin, one determines (as an average of a set
of many realizations) the survival probability $P(t)$, mean number of active sites $n(t)$,
and mean-square spread, $R^2 (t) = \langle \sum_j x_j(t)^2 \rangle/n(t)$.  (Here the sum is
over all active sites, with $x_j$ the position of the $j$-th active site, and the brackets
denote an average over all realizations.)
The spreading process is followed up to a certain maximum time $t_m$, with the system
taken large enough that activity does not reach the boundaries up to this time.
The expected scaling
behaviors at the critical point are conventionally denoted
$P(t) \sim t^{-\delta}$, $n(t) \sim t^\eta$ and $R^2 (t) \sim t^{z_{s}}$.  Away from the
critical point, these quantities show deviations from power laws.  $P(t)$, for example,
decays exponentially for $\lambda < \lambda_c$, and approaches a nonzero asymptotic value,
$P_\infty$, for $\lambda > \lambda_c$.

\begin{figure}[h]
\epsfysize=11cm \epsfxsize=14.5cm \centerline{ \epsfbox{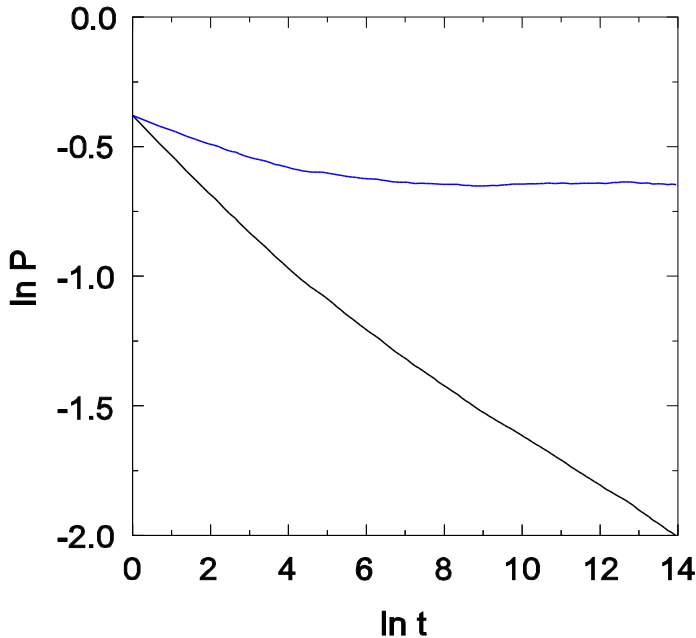}}
\caption{\footnotesize{Lower curve: survival probability $P$ versus time for
$v=0.1$ and $D=2$, $\lambda = 3.915$,
average over $10^4$ realizations. Upper curve:
Scaled survival probability $t^{\delta} P(t)$ using $\delta = 0.0971$.
}}
\label{ps3915}
\end{figure}

\begin{figure}[h]
\epsfysize=11cm \epsfxsize=14.5cm \centerline{ \epsfbox{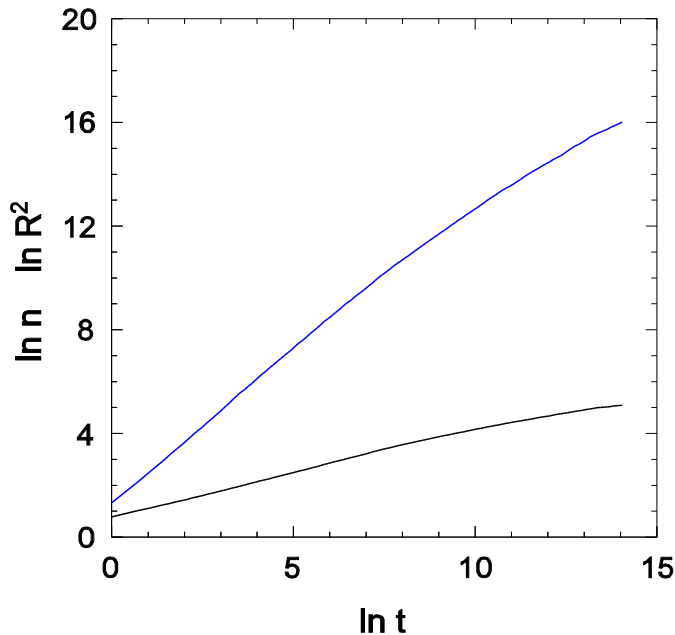}}
\caption{\footnotesize{Mean number $n$ of active sites (lower curve)
and mean-square distance from origin $R^2$ (upper curve) versus time;
parameters as in Fig. \ref{ps3915}.
}}
\label{nr2d2}
\end{figure}

In the CP with mobile disorder, this scaling picture appears to hold
for the survival probability but not for $n(t)$ and $R^2(t)$.
For example, Fig. \ref{ps3915} shows the decay of the survival probability for $v=0.1$
and $D=2$; $P(t)$ is well fit by an asymptotic power law with $\delta = 0.0971(5)$.
(Note however that at shorter times ($t < 100$) the decay is governed by a larger
exponent, with a value of about 0.147.)
The small differences between estimates for $\delta$ characterizing the decay of the
order parameter in CS, and that of the survival probability in spreading simulations
($\delta_c$ and $\delta_s$, respectively, in Table I), can be attributed to the
deviations from a pure power law noted above in the CS.
(Note also that the uncertainties reported for $\delta_s$ do not include any contribution
due to the uncertainty in $\lambda_c$, and so are smaller than the uncertainties in $\delta_c$.)
The behaviors of $n$ and $R^2$ at the critical point are
illustrated in Fig. \ref{nr2d2}.  While I have not found a simple functional form
capable of fitting these data,
it is clear that the growth is slower than a power law.  At short times $n$ and $R^2$
follow apparent power laws, with exponents of 0.35 and 1.22, respectively,
which are comparable to the DP spreading exponents $\eta \simeq 0.314$ and $z_{s} \simeq 1.27$.

\subsection{Critical vacancy concentration}

\begin{figure}[h]
\epsfysize=11cm \epsfxsize=14.5cm \centerline{ \epsfbox{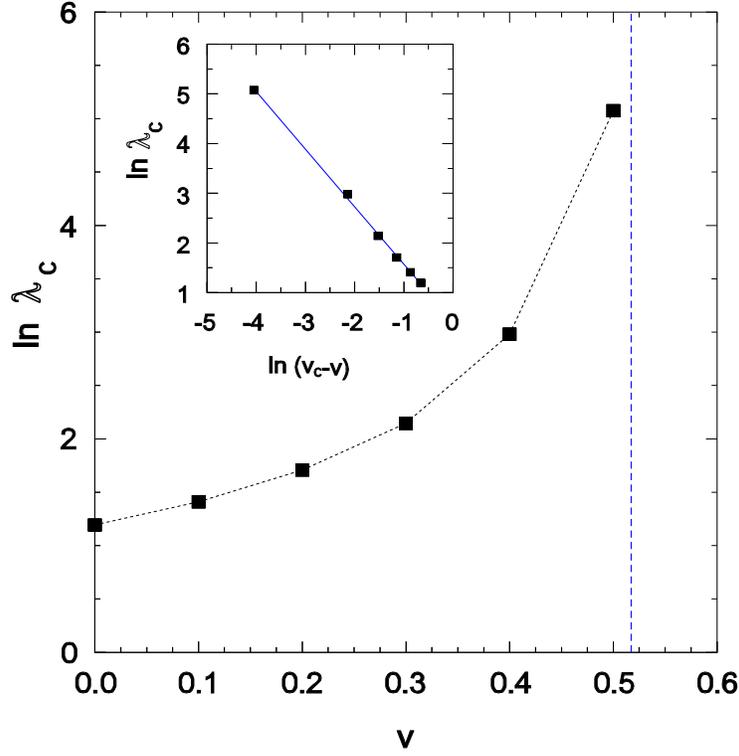}}
\caption{\footnotesize{Critical creation rate $\lambda_c$ versus vacancy
concentration $v$ for $D=1$  The dashed vertical line marks $v_c$.
Inset: the same data plotted versus $v_c - v$; the slope of the straight line
is -1.16.
}}
\label{lcvva}
\end{figure}

At a fixed diffusion rate, one expects $\lambda_c$ to be an increasing function of
$v$, as noted in Sec.~II; simulation results for $D=1$ confirm this.
(For $v \geq 0.2$ I determine $\lambda_c$ in conventional simulations, using the
criteria discussed above, in studies of systems with $L \leq 800$.)
To determine the critical vacancy concentration $v_c(D)$, at which $\lambda_c$
diverges, I perform conventional simulations
at an effectively infinite creation rate.  This is done by (1) allowing only {\it isolated}
active sites to become inactive, at a rate of unity, and (2) activating any nondiluted
site the instant it gains an active neighbor.  (Thus a string of inactive
sites are activated simultaneously when the right- or leftmost site acquires an active neighbor
via vacancy motion.)  These studies yield
$v_c = 0.517(1)$ for $D=1$, and $v_c \simeq 0.4182(5)$ for $D=0.2$.
Fig. \ref{lcvva} shows $\lambda_c (v)$ for $D=1$; the data suggest
that $\lambda_c \sim (v_c - v)^{-1.16}$ (see inset).
The available
results are consistent with $\beta/\nu_\perp$, and the critical moment ratio $m_c$,
being constant along the critical line for fixed $D=1$.

\begin{figure}[h]
\epsfysize=11cm \epsfxsize=14.5cm \centerline{ \epsfbox{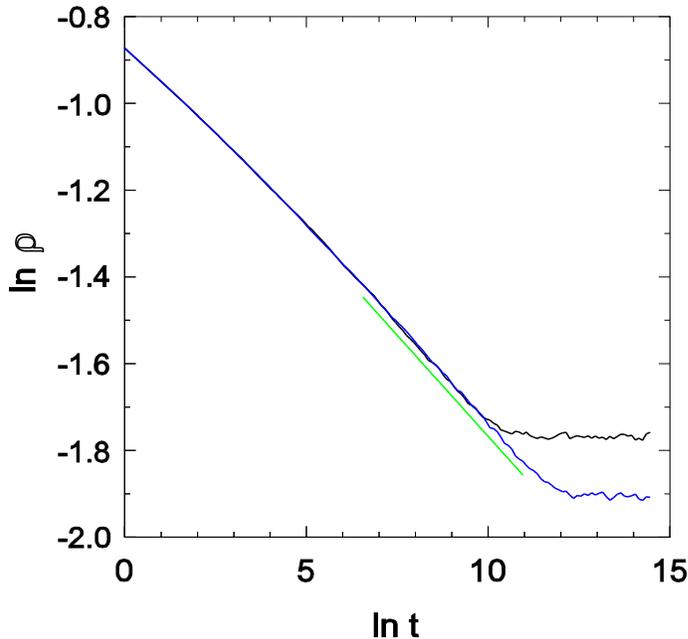}}
\caption{\footnotesize{Order parameter versus time for
$v=0.5176$, $\lambda = \infty$, with $D=1$.  System sizes $L=796$ (upper)
and 1592 (lower).  The slope of the
straight line is -0.093.
}}
\label{rt176b}
\end{figure}

\begin{figure}[h]
\epsfysize=11cm \epsfxsize=14.5cm \centerline{ \epsfbox{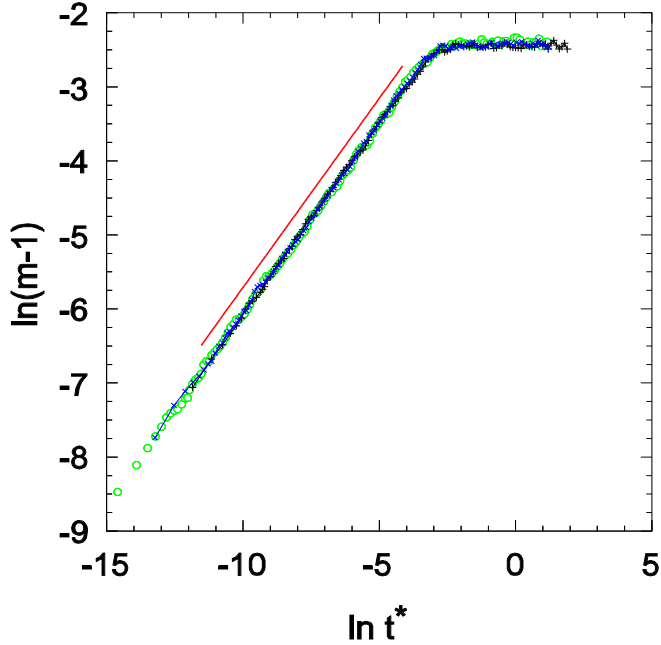}}
\caption{\footnotesize{Scaling plot of $m-1$ versus $t^* = t/L^z$ using
$z=1.98$.  Parameters
$v=0.5176$, $\lambda = \infty$, and $D=1$.  System sizes $L=398$ (+), $L=796$ ($\times$),
and 1592 (circles).  The slope of the straight line is 0.51.
}}
\label{mm176b}
\end{figure}

\begin{figure}[h]
\epsfysize=11cm \epsfxsize=14.5cm \centerline{ \epsfbox{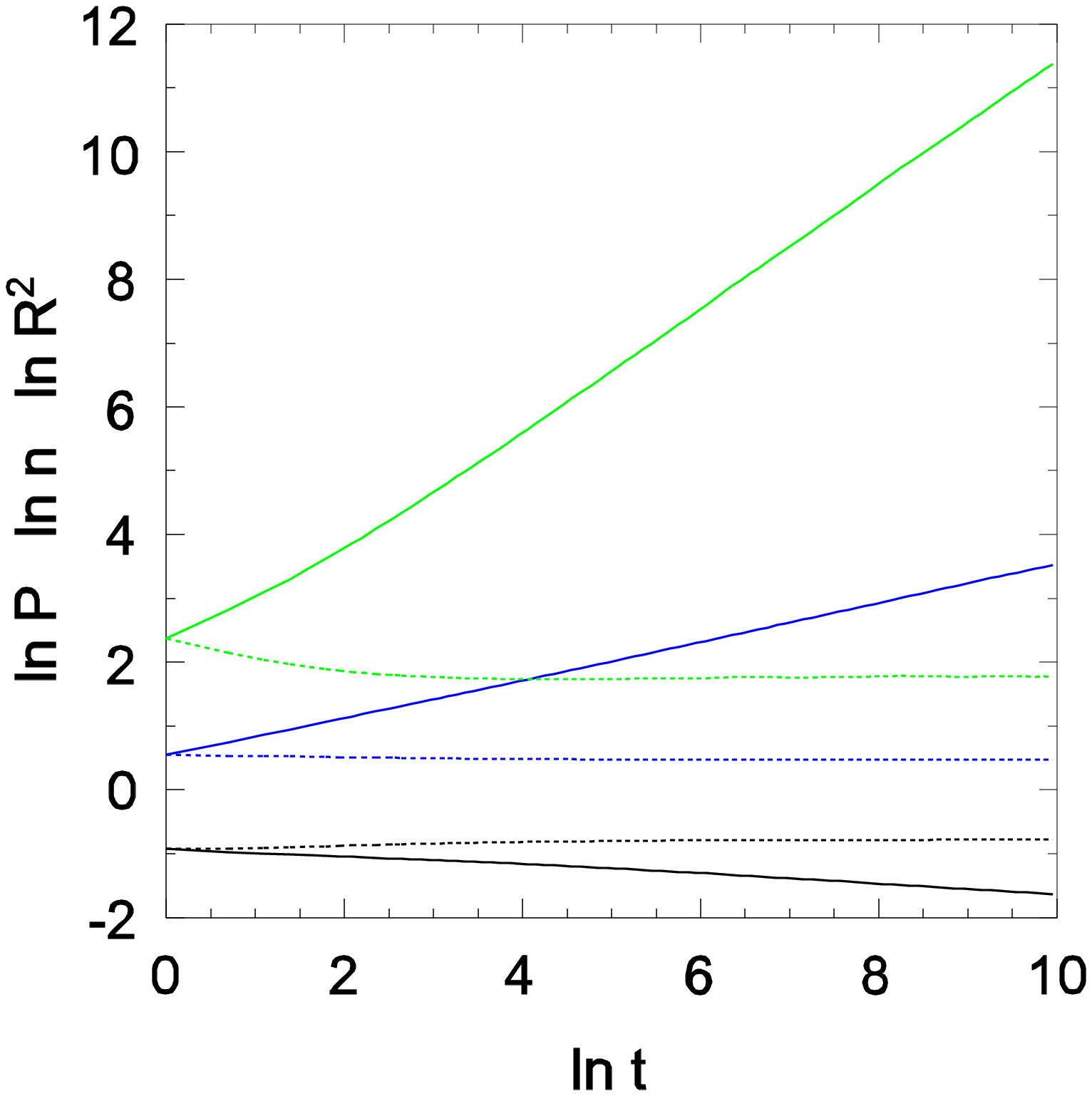}}
\caption{\footnotesize{Spreading simulations at the critical vacancy density.
Solid lines (lower to upper): $P(t)$, $n(t)$ and $R^2 (t)$.  The corresponding
dotted lines show the scaled quantities $P^* = t^\delta P$, $n^* = t^{-\eta} n$, and
$R^{*2} = t^{-z_s} R^2$, using $\delta = 0.086$, $\eta = 0.307$, and $z_s = 0.965$. Parameters
$v=0.516$, $\lambda = \infty$, $D=1$, system size $L=6000$.
}}
\label{sps516}
\end{figure}

I turn now to the scaling behavior at the critical vacancy density $v=v_c$, $\lambda = \infty$.
For $D=1$, finite-size scaling analysis of the order parameter yields $\beta/\nu_\perp = 0.184(20)$.
Unlike in the case $v=0.1$ discussed above, here the survival time follows a power-law with
$z = 1.98(3)$, close to the value for $z$ obtained via analysis of the growth of $m(t)$.
The short-time scaling behavior $\rho \sim t^{-\delta}$ is followed quite clearly
(see Fig. \ref{rt176b}), without the strong crossover effects observed for $v=0.1$ (Fig.
\ref{rt099}).  Fig. \ref{mm176b} is a scaling plot of $m-1$, using $t^* = t/L^z$, showing
a near-perfect data collapse for three system sizes, and a clean power law,
$m - 1 \sim t^{1/z}$.  There is also good evidence for power-law scaling
of the survival probability {\it and} of $n$ and $R^2$ at $v_c$, as shown in Fig. \ref{sps516}.
Thus the scaling behavior at the critical vacancy density
appears to be simpler than that observed for $v=0.1$.
Results for critical parameters at the critical vacancy density are summarized
in Table II.  For the two diffusion rates studied, the exponents and $m_c$ agree to within
uncertainty.
The spreading exponents satisfy the hyperscaling relation \cite{torre}
$4 \delta + 2 \eta = dz$ to within uncertainty.
The principal inconsistency in these data is in regard to the scaling relation $z = 2/z_s$.
In fact $2/z_s$ is about 8\% (5\%) greater than $z_m$ for $D=0.2$ ($D=1$).  Studies of larger
systems and a more precise determination of $v_c$ should help to resolve this discrepancy.

\section{Discussion}

I study the one-dimensional contact process with mobile vacancies, verifying that
the model exhibits a continuous phase transition between the active and the absorbing
phases, as in the undiluted model.  The scaling behavior is, however, considerably more
complex than the Fisher renormalization scenario that might be expected
by analogy with equilibrium systems with annealed disorder.  (The analogy, as noted,
is somewhat flawed since the
disorder variables are not in equilibrium with the other degrees of freedom.)
Cluster approximations appear to be
incapable of providing an accurate phase diagram.
Quantitative results on critical behavior are obtained using
conventional simulations (CS) and spreading studies.  Quasistationary
simulations yield results consistent with CS, except for a slightly smaller survival
time.

The scaling behavior is anomalous in several aspects.  First, the growth of the lifetime
$\tau$ at the critical point appears to be slower than a power law, as is also the case
for mean number of active sites and mean-square spread in spreading simulations
at the critical point.
The latter finding is reminiscent of the one-dimensional CP with {\it quenched} disorder
in the form of an annihilation rate randomly taking one of two values, independently at each
site \cite{bramson,webman}. In this case an intermediate phase arises,
in which the survival probability, starting from a
single active site, attains a nonzero value as $t \to \infty$, but the active region grows
in a sublinear manner with time.
The spreading simulations also show a crossover from scaling with exponent values
similar to those of DP at short times, to a different behavior at longer times.

In Sec. II it was argued that $\lambda_c \to \infty$ as $D \to 0$, and that $\lambda_c \to
\lambda_{c, pure}/(1-v)$ in the opposite limit.  The available data are
consistent with this (see Fig. \ref{lcv01});
extrapolation of the data for $v=0.1$ yields a limiting
($D \to \infty$) value of $\lambda_c = 3.65$, quite near $\lambda_{c,pure}/(1-v) \simeq 3.664$.

For vacancy density $v=0.1$, the critical exponents $\delta$ and $z$,
the ratio $\beta/\nu_\perp$, and moment ratio $m_c$ vary systematically with the vacancy diffusion
rate (Table I).  With increasing $D$ these parameters tend to their directed percolation values,
as might be expected, since the limit $D \to \infty$ corresponds to the usual
contact process, with a renormalized creation rate.  There is preliminary evidence that
some critical properties are insensitive to varying the vacancy concentration $v$ at a
fixed diffusion rate.
Preliminary results for critical behavior at the critical vacancy density $v_c$ suggest
a simpler scaling behavior, without strong crossover effects, at this point.
The static, transient, and spreading behavior is quite analogous to that observed
in the basic contact process, but with a different set of critical exponents.
This motivates the conjecture that for $v>0$
all points along the critical line $\lambda_c(v)$ flow, in the renormalization group
sense, to the same fixed point.  For smaller
values of $v$ one would then expect to observe a crossover between DP scaling and that
of the new fixed point, associated with mobile disorder.
More extensive studies are needed to verify and fill out this picture.

The effects of mobile reported here are more extensive than those found by Evron
et al. in a related model (EKS) \cite{evron}.  These authors do not find, for example,
changes in the static critical exponents.
In comparing the two models, it should be noted that the disorder in the CPMV is stronger than
that involved in the EKS model.  The difference nevertheless raises an interesting puzzle regarding
the universality of disorder effects.

Intuitively, the anomalous scaling properties of the CP with mobile disorder
are associated with fluctuations in the local, time-dependent
concentration of vacancies, which relax slowly in the long-wavelength limit.
Developing a quantitative theory of how
such fluctuations change the critical behavior is a challenge for
future work, with potential applications in epidemiology and the dynamics of
spatially distributed populations.  Investigation of the two-dimensional case
is also relevant for applications, and of intrinsic interest for developing a
more complete understanding of scaling.

\vspace{2em}

\noindent{\bf Acknowledgments}
\vspace{1em}

I thank Thomas Vojta, Jos\'e A. Hoyos, and Miguel A. Mu\~noz for
helpful comments.
This work was supported by CNPq and Fapemig, Brazil.

\newpage

\bibliographystyle{apsrev}

\begin{thebibliography}{100}


\bibitem{marro}
        J. Marro and R. Dickman,
        {\it Nonequilibrium Phase Transitions in Lattice Models}
        (Cambridge University Press, Cambridge, 1999).

\bibitem{hinrichsen}
        H. Hinrichsen,
        Adv. Phys. {\bf 49} 815 (2000).

\bibitem{granada}
        M. A. Mu\~noz, R. Dickman, R. Pastor-Satorras, A. Vespignani,
        and S. Zapperi, in {\it Procedings of the Sixth Granada Seminar
        on Computational Physics}, edited by J. Marro and P. L. Garrido
        (AIP, New York, 2001).

\bibitem{lubeck04}
        S. L\"ubeck,
        Int. J. Mod. Phys. B {\bf 18}, 3977 (2004).

\bibitem{odor04}
        G. \'Odor,
        Rev. Mod. Phys {\bf 76},  663 (2004).

\bibitem{harris}
        T. E. Harris,
        Ann. Probab. {\bf 2}, 969 (1974).

\bibitem{takeuchi}
        K. A. Takeuchi, M. Kuroda, H. Chat\'e, and M. Sano,
        Phys. Rev. Lett. {\bf 99}, 234503 (2007).

\bibitem{janssen}
        H.-K. Janssen,
        Z. Phys. B {\bf 42}, 151 (1981).

\bibitem{grassberger}
        P. Grassberger,
        Z. Phys. B {\bf 47}, 465 (1982).

\bibitem{iwandcp}
        I. Jensen and R. Dickman,
        J. Phys. A{\bf 26}, L151 (1993).

\bibitem{noest}
        A. J. Noest,
        Phys. Rev. Lett. {\bf 57}, 90 (1986);
        Phys. Rev. B{\bf 38}, 2715 (1988).

\bibitem{agrd}
        A. G. Moreira and R. Dickman,
        Phys. Rev. E{\bf 54}, R3090 (1996);
        R. Dickman and A. G. Moreira,
        Phys. Rev. E{\bf 57}, 1263 (1998).

\bibitem{janssen97}
        H. K Janssen,
        Phys. Rev. E{\bf 55}, 6253 (1997).

\bibitem{bramson}
        M. Bramson, R. Durrett, and R. H. Schonmann,
        Ann. Prob. {\bf 19}, 960 (1991).

\bibitem{webman}
        I. Webman, D. ben-Avraham, A. Cohen, and S. Havlin,
        Phil. Mag. B {\bf 77}, 1401 (1998).

\bibitem{cafiero}
        R. Cafiero, A. Gabrielli, and M. A. Mu\~noz,
        Phys. Rev. E {\bf 57}, 5060 (1998).

\bibitem{hooyberghs}
        J. Hooyberghs, F. Igloi, and C. Vanderzande,
        Phys. Rev. Lett. {\bf 90}, 100601 (2003);
        Phys. Rev. E {\bf 69}, 066140 (2004).

\bibitem{vojta}
        T. Vojta and M. Dickison,
        Phys. Rev. E {\bf 72}, 036126 (2005).

\bibitem{hoyos}
        J. A. Hoyos,
        Phys. Rev. E {\bf 78}, 032101.

\bibitem{deoliveira08}
        M. M. de Oliveira and S. C. Ferreira,
        J. Stat. Mech. (2008) P11001.

\bibitem{evron}
        G. Evron, D. A. Kessler, and N. M. Shnerb,
        eprint: arXiv:0808.0592.

\bibitem{fisherren}
        M. E. Fisher,
        Phys. Rev. {\bf 176}, 257 (1968).

\bibitem{harris74}
        A. B. Harris,
        J. Phys. C {\bf 7}, 1671 (1974).

\bibitem{berker}
        A. N. Berker,
        Physica A {\bf 194}, 72 (1993).

\bibitem{alonso-munoz}
        J. J. Alonso and M. A. Mu\~noz,
        Europhys. Lett. {\bf 56}, 485 (2001).

\bibitem{cardy}
        J. L. Cardy and R. L. Sugar,
        J. Phys. A {\bf 13}, L423 (1980).

\bibitem{hinrichsenbjp}
        H. Hinrichsen,
        Braz. J. Phys. {\bf 30}, 69 (2000).

\bibitem{dep}
        R. Kree, B. Schaub, and B. Schmittmann,
        Phys. Rev. A {\bf 39}, 2214 (1989).

\bibitem{benav}
        D. ben-Avraham and J. Köhler,
        Phys. Rev. A {\bf 45}, 8358 (1992).

\bibitem{mancam}
        R. Dickman,
        Phys. Rev. E {\bf 66}, 036122 (2002).

\bibitem{silva}
        R. da Silva, R. Dickman, and J. R. Drugowich de Felício
        Phys. Rev. E {\bf 70}, 067701 (2004).

\bibitem{dic-jaff}
        R. Dickman and J. K. Leal da Silva,
        Phys. Rev. E, {\bf 58} 4266 (1998).

\bibitem{qssim}
        M. M. de Oliveira and R. Dickman,
        Phys. Rev. E {\bf 71}, 016129 (2005).

\bibitem{epidif2}
        R. Dickman and D. S. Maia,
        J. Phys. A {\bf 41}, 405002 (2008).

\bibitem{torre}
        P. Grassberger and A. de la Torre,
        Ann. Phys. (N.Y.) {\bf 122}, 373 (1979).

\bibitem{jensen96}
        I. Jensen,
        J. Phys. A {\bf 29}, 7013 (1996).

\end{thebibliography}

\newpage

{\sf Table I. Critical parameters of the CPMV with
vacancy density $v=0.1$. $\delta_c$
and $\delta_s$ denote, respectively, the exponent as determined in conventional
and in spreading simulations. DP values from \cite{jensen96}.} \vspace{1em}

\begin{center}
\begin{tabular}{c|c c c c c c}
\hline \hline
$D$ & $\lambda_c$ & $\beta/\nu_\perp$ & $m_c$    &   $z$   & $\delta_c$ & $\delta_s$ \\
\hline

0.5 &  4.375(2)  &     0.175(3)    & 1.076(2)   & 2.65(4) & 0.0745(15) & 0.0765(2) \\

1   &  4.099(1)  &     0.191(3)    & 1.085(2)   & 2.49(1) & 0.086(2)   & 0.0837(5)\\

2   &  3.915(1)  &     0.205(3)    & 1.096(3)   & 2.36(5) & 0.101(4)   & 0.0971(5)\\

5   & 3.7746(10) &     0.235(4)    & 1.123(4)   & 1.92(2) & 0.137(3)   & 0.1293(5)\\

\hline \hline
DP &  3.29785(2) &     0.2521(1)   & 1.1736(1)  & 1.58074(4) & 0.15947(3) & (=$\delta_c$) \\

\hline \hline
\end{tabular}
\end{center}
\vspace{3em}

{\sf Table II. Critical parameters of the CPMV at the critical
vacancy density $v_c$. $\delta_c$
and $\delta_s$ as in Table I. $z_m$ is the dynamic exponent as determined from the
growth of $m$ in conventional simulations. DP values from \cite{jensen96}.} \vspace{1em}

\begin{center}
\begin{tabular}{c|c c c c c c c c}
\hline \hline
$D$ & $v_c$ & $\beta/\nu_\perp$ & $m_c$    & $z_m$ & $\delta_c$ & $\delta_s$ & $\eta$ & $z_s$ \\
\hline

0.2 & 0.4182(5) &  0.174(6)     & 1.083(3) & 1.95(4) & 0.087(2) & 0.086(2) & 0.303(3) & 0.95(1) \\

1   & 0.517(1)  &  0.184(20)    & 1.084(11)& 1.98(3) & 0.091(4) & 0.086(2) & 0.307(1) & 0.965(10) \\

\hline \hline
DP  &  -        &  0.2521(1)    & 1.1736(1)& 1.58074(4)& 0.15947(3) & (=$\delta_c$) & 0.31368(4)
& 1.26523(3) \\

\hline \hline
\end{tabular}
\end{center}
\vspace{5em}

\newpage
\noindent {\bf Appendix: Pair approximation}
\vspace{1em}

It is straightforward to construct the pair approximation to the CPMV.  For convenience I denote
vacant, inactive, and active sites by $v$, 0, and 1, respectively.
In this approximation the dynamical variables
are the probabilities $p(00) \equiv (00)$, $(01)$, etc.  Note that since vacancies
hop independently of the states (0 or 1) of the nonvacant sites, $(vv) = v^2$ in the stationary
state, and during the entire evolution, if this equality holds initially.
Since $v = (vv) + (0v) + (1v)$, we then have $(1v) = v - v^2 - (0v)$, leaving four independent pair
probabilities.  (The normalization condition, $(00) + (11) + (vv) +2[(01) + (0v) + (1v)]$,
permits to eliminate one further variable.)

Consider, for example, the transition $00 \to 01$.  It occurs at rate
$\lambda/2$ provided the site to the right of the pair bears a particle, so the contribution to
$d(00)/dt$ due to this event is $-(\lambda/2) (001)$.  (The mirror event makes an identical
contribution.)  In the pair approximation, three-site probabilities are written in terms
of the one- and two-site probabilities, for example, $(001) \simeq (00)(01)/(0)$.
Proceeding in this manner we obtain

\begin{equation}
\frac{d(00)}{dt} = 2(01) + D \frac{(0v)^2}{v} - \frac{(00)}{(0)}[\lambda (01) + D(0v)]
\label{d00dt}
\end{equation}

\begin{equation}
\frac{d(01)}{dt} = (11) + \frac{\lambda}{2}\frac{(00)(01)}{(0)} + D \frac{(0v)(1v)}{v}
- (01)\left[1 + \frac{D}{2} \left( \frac{(1v)}{(1)} + \frac{(0v)}{(0)} \right)
+ \frac{\lambda}{2} \left(1 + \frac{(10)}{(0)}\right)  \right]
\label{d01dt}
\end{equation}

\begin{eqnarray}
\frac{d(0v)}{dt} &=& (1v) + \frac{D}{2}\left[\frac{(00)(0v)}{(0)} + v(0v) + \frac{(01)(1v)}{(1)} \right]
\nonumber
\\
&\;\;& - \frac{D}{2} (0v) \left[\frac{(0v)}{v} + \frac{(1v)}{v} + \frac{(0v)}{(0)} \right]
- \frac{\lambda}{2} (0v) \frac{(10)}{(0)}
\label{d0vdt}
\end{eqnarray}

\noindent and

\begin{equation}
\frac{d(11)}{dt} = \lambda (01) \left(1 + \frac{(10)}{(0)} \right)
 + D \frac{(1v)^2}{v} - (11)\left( 2 + D \frac{(1v)}{(1)} \right)
\label{d11dt}
\end{equation}
\vspace{1em}

\noindent Integrating these equations numerically, one can determine the critical creation rate $\lambda_c$
as a function of $v$ and $D$.

\end{document}